\documentclass[apjl]{emulateapj}
\usepackage{hyperref}
\usepackage{booktabs}

\usepackage{natbib}

\usepackage{color}

\newcommand{\software}[1] {\textit{#1}}

\newcommand{\gwevent} {GW150914}

\shorttitle{INTEGRAL gamma-ray upper limit on \gwevent }
\shortauthors{Savchenko et al.}

\begin{document}

\title{INTEGRAL upper limits on gamma-ray emission associated with
the gravitational wave event \gwevent}

\author{V.~Savchenko$^{1}$, C.~Ferrigno$^{2}$, S.~Mereghetti$^{3}$,L.~Natalucci$^{4}$, A.~Bazzano$^{4}$, E.~Bozzo$^{2}$, S.~Brandt$^{5}$, T.~J.-L.~Courvoisier$^{2}$, R.~Diehl$^{6}$, L.~Hanlon$^{7}$, A.~von~Kienlin$^{6}$, E.~Kuulkers$^{8}$, P.~Laurent$^{9,10}$, F.~Lebrun$^{9}$, J.~P.~Roques$^{11}$, P.~Ubertini$^{4}$, G.~Weidenspointner$^{6,12}$}

\affil{$^{1}$Fran\c{c}ois Arago Centre, APC, Universit\'e Paris Diderot, CNRS/IN2P3, CEA/Irfu, Observatoire de Paris, \\
Sorbonne Paris Cit\'e, 10 rue Alice Domon et L\'eonie Duquet, 75205 Paris Cedex 13, France\\
$^{2}$ISDC, Department of astronomy, University of Geneva, chemin d'\'Ecogia, 16 CH-1290 Versoix, Switzerland\\
$^{3}$INAF, IASF-Milano, via E.Bassini 15, I-20133 Milano, Italy \\
$^{4}$INAF-Institute for Space Astrophysics and Planetology, Via Fosso del Cavaliere 100, 00133-Rome, Italy\\
$^{5}$DTU Space - National Space Institute Elektrovej - Building 327 DK-2800 Kongens Lyngby Denmark\\
$^{6}$Max-Planck-Institut f\"{u}r Extraterrestrische Physik, Garching, Germany\\
$^{7}$Space Science Group, School of Physics, University College Dublin, Belfield, Dublin 4, Ireland\\
$^{8}$European Space Astronomy Centre (ESA/ESAC), Science Operations Department 28691, Villanueva de la Ca\~nada, Madrid, Spain\\
$^{9}$APC, AstroParticule et Cosmologie, Universit\'e Paris Diderot, CNRS/IN2P3, CEA/Irfu, Observatoire de Paris, \\
Sorbonne Paris Cit\'e, 10 rue Alice Domont et L\'eonie Duquet, 75205 Paris Cedex 13, France.\\
$^{10}$DSM/Irfu/Service d'Astrophysique, Bat. 709 Orme des Merisiers CEA Saclay, 91191 Gif-sur-Yvette Cedex, France\\
$^{11}$Universit\'e Toulouse; UPS-OMP; CNRS; IRAP; 9 Av. Roche, BP 44346, F-31028 Toulouse, France\\
$^{12}$European XFEL GmbH, Albert-Einstein-Ring 19, 22761, Hamburg, Germany\\
}

\begin{abstract}
Using observations of the INTErnational Gamma-Ray Astrophysics
Laboratory (INTEGRAL), we put upper limits on the gamma-ray and hard
X-ray prompt emission associated with the gravitational wave event
\gwevent, discovered by the LIGO/Virgo collaboration.  The
omni-directional view of the INTEGRAL/SPI-ACS has allowed us to
constrain the fraction of energy emitted in the hard X-ray
electromagnetic component for the full high-probability sky region of
LIGO trigger. Our upper limits on the hard X-ray fluence at the
time of the event range from $F_{\gamma}=2 \times
10^{-8}$~erg~cm$^{-2}$ to $F_{\gamma}=10^{-6}$~erg~cm$^{-2}$ in the
75~keV~-~2~MeV energy range for typical spectral models.  Our results
constrain the ratio of the energy promptly released in gamma-rays in
the direction of the observer to the gravitational wave energy
E$_\gamma/$E$_\mathrm{GW}<10^{-6}$.  We discuss the implication of
gamma-ray limits on the characteristics of the gravitational wave
source, based on the available predictions for prompt electromagnetic
emission.
\end{abstract}

\section{Introduction}

Gravitational waves were predicted nearly one hundred years ago as a
natural consequence of general relativity \citep{relativity}, but up
to now only indirect evidence of their existence has been found by
measuring the time evolution of orbital parameters of binary pulsars
\citep{gw_indirect,kramer2006}. The direct detection of gravitational
waves is challenging since it relies on measurements of the relative
change in distance of the order of $10^{-22}$. This will be achieved,
for low frequency signals ($10^{-4}$--1 Hz), with the space-based
eLISA mission to be launched after 2030 \citep{elisa}, while it is
currently possible at higher frequency (10--$10^4$\,Hz), thanks to the
ground-based advanced LIGO \citep{aligo} and Virgo \citep{avirgo}
detectors.  Advanced LIGO has been in operation since September 2015
with the first science run extending to January 2016 and a sensitivity
enabling routine detection of gravitational waves from merging compact
binaries. Once a possible trigger has been recorded, it is vital to
conduct multiwavelength observations to search for additional
information about this event. The LIGO/Virgo collaboration recently
reported the first gravitational-wave event, \gwevent, detected on
2015-09-14 at 09:50:45 UTC, with a false alarm probabilty of  
less than one event per 203\,000 years \citep{triggerpaper,triggerparampaper}.  
Here, we exploit the data
obtained by the INTEGRAL satellite \citep{integral}, which was fully
operational at the time of the gravitational-wave trigger, to derive
limits on the hard X-ray and gamma-ray emission associated with this
event.

\section{INTEGRAL/SPI-ACS}

The SPI instrument onboard INTEGRAL \citep{spi} comprises an active
anti-coincidence shield \citep[ACS,][]{spiacs} made of 91 BGO (Bismuth
Germanate, Bi$_4$Ge$_3$O$_{12}$) scintillator crystals\footnote{only
  89 are currently functional}.  Besides its main function of
shielding the SPI germanium detectors, the ACS is also used as a
nearly onmidirectional detector of transient events with a large
effective area (up to 1~m$^2$) at energies above $\sim$75~keV
\citep{spiacs}.
The ACS data consist of event rates integrated over all the
scintillator crystals with a time resolution of 50 ms. The typical
number of counts per 50 ms time bin ranges from about 3000 to 6000 (or
more during periods of high Solar activity). Since only a single
integrated rate is recorded for the whole instrument, no spectral and
directional information is available.
Contrary to most instruments for the detection of GRBs, the ACS read-out does not rely on any trigger.
Thus a complete time history of the detector count rate is continuously recorded for $\sim$90\% of the time\footnote{instruments are switched-off near perigee of every revolution;  until January 2015, the
INTEGRAL orbit lasted three sidereal days. Afterwards, it was reduced to 2.7 sidereal days to allow for a safe satellite disposal in 2029. } and simultaneously covering nearly the whole sky.

SPI is partially surrounded by the satellite structure and by the
other INTEGRAL instruments, which, by shielding the incoming photons,
affect the response of the ACS in the different directions.
Therefore, the ACS response must be determined through detailed
simulations which take into account the whole satellite structure.  We
developed a \software{GEANT3} Monte-Carlo model based on the INTEGRAL
mass model \citep{sturner03} and simulated the propagation of
monochromatic parallel beams of photons in the 50 keV to 100 MeV
range. For each energy we simulated 3072 sky positions (16-side
\software{HEALPix}\footnote{http://healpix.sourceforge.net}
grid). This enables us to generate an instrumental response function
for any sky position, which can then be used to compute the expected
number of counts for a given intrinsic source spectrum. We have
verified that the response produces valid results for the bursts
detected simultaneously by SPI-ACS and other detectors, primarily
Fermi/GBM, with an accuracy better than 20\%.

\section{Results}

SPI-ACS was operating nominally at the time of the LIGO trigger on
2015-09-14 at 09:50:45 UTC, yielding an uninterrupted count rate from
33 hours before to 19 hours after the event. The background was
relatively stable and low, with a rate of $\sim 7 \times
10^4~$counts/s. The main limit to the sensitivity is set by the
Poisson noise in the background rate. In addition to the high-count
rate approximation of the Poisson process, there is an excess variance
which changes from 3\% to 10\% on a time scale of the order of one
year, and increases in case of strong solar activity. This excess
noise is related to multiple events in the detector and to the solar
activity. The total noise at every time scale can still be
well-described by a gaussian process \citep{savchenko12}. We measure
the average background and its variance in the vicinity of the region
of interest, from $-$1000~s to +1000~s from the trigger and use it in
the computation of significance and upper limits.

We investigated the light curve at $-$30 to +30 s from the trigger
time on 5 time scales from 0.05 to 10 s. These time scales correspond
to the expected accretion time scales in the compact binary
coalescence \citep{lee07}. We do not detect any obvious signal
coincident with the GW trigger. We derived a maximum post-trial peak
significance of $\sim$0.5$\sigma$ with a time scale of 1 s, at 26.4 s
after the GW trigger. Such an excess is clearly not significant.

\begin{figure}
  \centering 
  \vspace{0.5cm}
  \includegraphics[width=0.75\columnwidth, angle =-90]{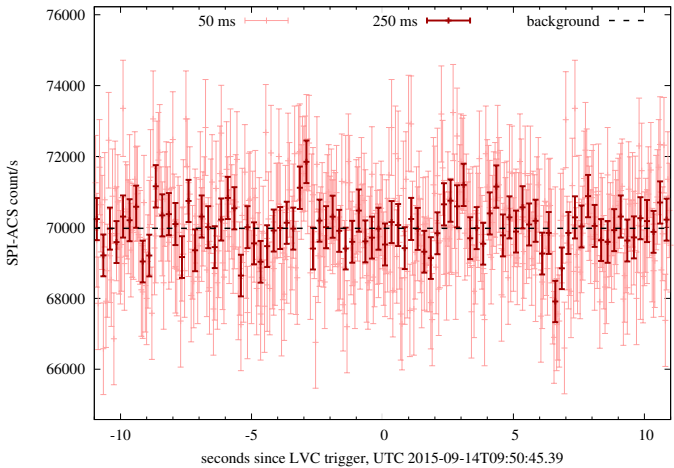}
  \caption{INTEGRAL/SPI-ACS lightcurve in $\pm$10~s around
    \gwevent\ trigger time.  Light red symbols represent the
    measurements at the natural instrument time resolution of 50 ms;
    dark red points are rebinned to 250 ms. The dashed black curve is
    the background level estimated from a long-term average.}
  \label{fig:lightcurve}
\end{figure}

A zoom on the light curve from $-$10 to +10 s around the trigger
time is shown in Figure~\ref{fig:lightcurve}. The excess at $T_0$-3~s,
where $T_0$ is the GW trigger time, is compatible with regular
background variability.  A similar, but negative, deviation can be seen at
$T_0+$7~s.

The upper limit on the total number of observed photons depends on the
assumed duration of the event. The results for different search time
scales are summarized in Table~\ref{tab:limits}. The dependence of the
upper limit on the burst duration remains the same for any sky position
or burst spectrum. In what follows we assume a typical duration for a
short GRB, 1~s.

\begin{table}[ht]
\centering
\caption{Three sigma upper limit on the possible gamma-ray counterpart fluence}
\begin{tabular}{ c c c c }

\toprule
Time scale & total counts & \multicolumn{2}{c}{fluence} \\
\vspace{0.3cm}
(seconds) & & \multicolumn{2}{c}{(erg cm$^{-2}$)} \\

& & best 95\% & worst 5\%  \\
\midrule

 10 & 4319 & $ 3.5-4.5 \times 10^{-7}$ & $ 1.1-1.4 \times 10^{-6}$
 \\ 1 & 1410 & $ 1.3-1.5 \times 10^{-7}$ & $ 3.7-4.7 \times 10^{-7}$
 \\ 0.25 & 727 & $ 5.8-7.6 \times 10^{-8}$ & $ 1.9-2.4 \times 10^{-7}$
 \\ 0.1 & 200 & $ 1.6-2.1 \times 10^{-8}$ & $ 5.2-6.6 \times 10^{-8}$
 \\ 0.05 & 220 & $ 1.8-2.3 \times 10^{-8}$ & $ 5.7-7.3 \times 10^{-8}$
 \\ \bottomrule
\vspace{0.3cm}
\end{tabular}
\label{tab:limits}
\tablecomments{The fluence range is calculated in the 75-2000~keV range,
  assuming two standard hard and soft GRB spectra, characterized
  by smoothly broken power law \citep[Band model;][]{band93} with parameters
  $\alpha=-0.5$, $\beta=-1.5$, $E_{peak}=1000~$keV and $\alpha=-1.5$,
  $\beta=-2.5$, $E_{peak}=500~$keV. Best sensitivity applies to 95\%
  of the trigger localization region, for the remaining 5\% we provide
  a less constraining limit.}
\end{table}

\begin{figure}
  \centering 
  \vspace{0.5cm}
   \includegraphics[width=0.9\columnwidth]{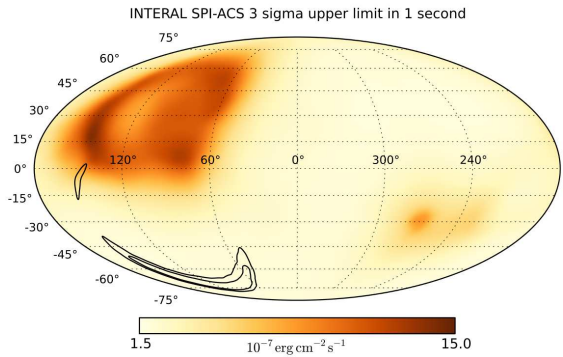}
  \caption{ INTEGRAL/SPI-ACS 3 sigma upper limit in 1 second for a
    characteristic short hard GRB spectrum: Band model with parameters
    $\alpha=-0.5$, $\beta=-2.5$, $E_{peak}=1000~$keV. In black
    contours regions (50\% and 90\%) we show the most accurate
    \gwevent\ trigger localization from the LALInference
    \citep{triggerempaper}.}
  \label{fig:sky}
\end{figure}
\begin{figure}
  \centering 
  \includegraphics[width=0.7\columnwidth, angle =-90]{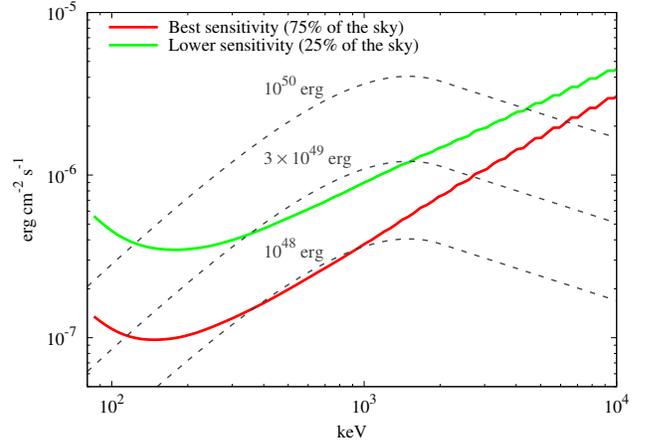}
  \caption{INTEGRAL/SPI-ACS 3-sigma sensitivity as function of energy
    averaged over each of the two sky regions, corresponding to
    optimal orthogonal orientation and the least favorable directions
    shaded by heavy satellite material. Dashed curves correspond to the
    hard GRB spectrum used in Fig.~\ref{fig:sky}, scaled to reproduce several values of the total
    energy released in the 75-2000~keV band, assuming a distance to the source of
    410~Mpc.}
  \label{fig:spectra}
\end{figure}

In order to put an upper limit on the signal fluence, we have to
investigate the different assumptions on the spectrum and sky
coordinates. Figure~\ref{fig:sky} shows the upper limit on the
75--2000~keV fluence in 1~second for a typical short hard GRB
spectrum: smoothly broken power law (Band model) with parameters
$\alpha=-0.5$, $\beta=-2.5$, $E_{peak}=1000~$keV
\citep{ghirlanda09}. SPI-ACS observed the full sky and in particular
covered about 95\% of the \gwevent\ localization confidence area with
a sensitivity at most 20\% lower than that reached in the most
favorable position. The weighted average of the limiting fluence in
this region is 4\% higher than that of the best, while it is a factor
3 less favorable over the remaining 5\% of the localization
region. The reduced sensitivity is caused by the opacity of the
satellite structure and the other INTEGRAL instruments. The limit
depends, however, on the incident spectrum: for harder spectra the
low-sensitivity regions are less pronounced. Figure~\ref{fig:spectra}
illustrates the energy dependency of the SPI-ACS sensitivity for two
sky regions. The low energy threshold of ACS around 75~keV limits our
low-energy sensitivity. At high energy, the effective area is
approximately constant, slowly increasing above ~1~MeV. A spectrum
typical for a hard gamma-ray burst (Band model with
parameters $\alpha=-0.5$, $\beta=-2.5$, $E_{peak}=1000~$keV) is scaled to reproduce different values
of the total energy release in the 75-2000~keV band, assuming a distance of 410~Mpc.

\subsection{IBIS results}

The IBIS instrument \citep{ubertini03} is composed of two detectors,
ISGRI \citep[20-1000\,keV;][]{lebrun03} and PICsIT
\citep[175\,keV--10\,MeV;][]{dicocco03}, which using a coded mask
provide images over a field of view of 30$^{\circ}\times30^{\circ}$.
The ISGRI data are used to automatically search and localize in real
time GRBs and other transient through the INTEGRAL Burst Alert System
\citep[IBAS;][]{mereghetti03}.  IBAS did not reveal any new transient
in the IBIS field of view at the time of the LIGO trigger,
down to a peak flux sensitivity of $\sim$0.1 ph cm$^{-2}$ s$^{-1}$
(20--200 keV, 1~s integration time).  We also carried out an off-line
search in the time interval 09:28 -- 10:00 UT, again with negative
results.  Note, however, that the instruments of INTEGRAL were pointed at a position
(R.A.=271$^{\circ}$, Dec.=--31$^{\circ}$) outside the high-probability
region of the gravitational signal. This prevented also the X-ray monitor
instrument JEM-X \citep{jemx} to collect constraining data.

IBIS can also provide response to photons outside the field of view,
due to high-energy photons passing through the passive and active
shields of the instrument, allowing the detection of transient events.
Indeed, most of the shielding of IBIS is passive and relatively thin,
becoming transparent to photons above $\sim$200~keV. For high energy
events, ratemeters of the PICsIT detector are available in 8 energy
bands in the range 210~keV~-~2600~keV.  We investigated the count rate
light curve in $\pm$10~s around the \gwevent\ for possible excesses on
time scales from 0.016~s to 10~s, but found no positive signal. We set
3-sigma upper limits to fluences in the 570~keV~-~1200~keV energy range of
2.5$\times 10^{-7}$~erg~cm$^{-2}$ and 6.5$\times
10^{-7}$~erg~cm$^{-2}$ assuming durations of 1~s and 10~s,
respectively. These values apply to a fully exposed detector area. The
detection efficiency is highly dependent on the source position and
it is considerably reduced 
for sources located at large angles with respect to the instrument
pointing direction, owing
to the lower exposed area and the presence of the 2~cm
thick BGO anticoincidence shield.  The localization region of
\gwevent\ is positioned at large offset ($\sim$80 to 140 degrees) from
the telescope axis. This implies that the sensitivity is decreased and
is strongly dependent on the source position in the sky. Nevertheless,
the PICsIT observation provides an important independent limit on
gamma-ray emission above 500~keV associated with the \gwevent.

\subsection{On the Fermi/GBM candidate}

The Fermi/GBM team reported a possible hard X-ray transient on
2015-09-14 at 09:50:45.8 UTC, about 0.4~s after the reported
LIGO burst trigger time, and lasting for about one second
\citep{gbmtrigger,gbmpaper}. The light travel time can introduce a
time difference between INTEGRAL and Fermi detections of up to
$\pm$0.5~s, depending on the source position within the LVC error
region. We do not observe any excess within a -0.5~s to +0.5~s window
around the Fermi/GBM trigger (Figure~\ref{fig:lightcurve}), and set a
3-sigma upper limit of $1.5\times 10^{-7}$~erg~cm$^{-2}$ for one-second integration time, assuming a typical short hard GRB,
characterized by Band model with parameters $\alpha=-0.5$,
$\beta=-2.5$, $E_{peak}=1000~$keV. A substantial part of the candidate
event in the GBM comes from the high-energy BGO detector, above
100~keV \citep{gbmtrigger}, where the Fermi/GBM effective area is
about a factor 30-40 smaller than that of the INTEGRAL/SPI-ACS.

\citet{gbmpaper} find that preferable localization of their candidate
is in the direction of the Earth, or close to it, limited to the
southern, dominant, arc of the \gwevent\ localization. Assuming the
preferred localization, they conclude that the spectrum can be best fit
by a hard powerlaw with a slope of $-1.4$ and 10-1000~keV fluence of
$2.4_{-1.0}^{+1.7} \times 10^{-7}$~erg~cm$^{-2}$. Extrapolating this
spectrum to the full 75~keV--100~MeV energy range accessible to the
SPI-ACS without a cutoff is clearly unphysical and incompatible with
the SPI-ACS upper limit. 
However, no best fit parameters for a model comprising a cutoff powerlaw
are reported.
On the other hand, \cite{gbmpaper} found a best fit to the Comptonized
model in the north-eastern tip of the the southern arc, with a powerlaw index
$\alpha^\mathrm{COMP}=-0.16$ and
$E_\mathrm{peak}^\mathrm{COMP}=3500~keV$, harder than a typical
Fermi/GBM spectrum. We assume this spectral model to compute the
expected signal in the SPI-ACS: for the southern (northern) arc,
SPI-ACS would detect 4740 (1650) counts, with a signal significance of
15 (5) sigma above the background.  It should be noticed that the
northern arc is disfavored by both the GBM and the LIGO
localizations.

We stress that to compare the GBM and SPI-ACS sensitivities,
it is inappropriate to use a soft spectral model as in the computation
of our early fluence upper limits \citep{acsgcn}, since the spectral
properties of the GBM candidate are very different. 

Considering the reported hardness of the GBM candidate, and the
favorable orientation of the SPI-ACS with respect to the
\gwevent\ localization, we are inclined to claim that the
non-detection by SPI-ACS disfavors a cosmic origin of the Fermi/GBM
excess. If the origin of the event was near the Earth, INTEGRAL would
not detect it, due to the large INTEGRAL - Earth distance at the time of
\gwevent\ (140\,000~km). \citet{gbmpaper} discussed a possible terrestrial
origin of the GBM excess, and came to the conclusion that it is not
compatible with the characteristics of a terrestrial gamma-ray
flash. However, they do not exclude a possibility that the event had
a magnetospheric origin. Eventually, considering that the false
alarm probability of the GBM association relatively high
\citep[0.2\%;][]{gbmpaper} and SPI-ACS does not detect it, 
it is likely that the GBM excess is a random background fluctuation.

\section{Discussion}

\subsection{Model-independent limit}

INTEGRAL/SPI-ACS is the only instrument covering the whole
\gwevent\ position error region at the time of the GW trigger. The
limit depends on the position, burst duration, and the assumed spectral model,
and ranges from $F_{\gamma}=2 \times 10^{-8}$~erg~cm$^{-2}$ to
$F_{\gamma}=10^{-6}$~erg~cm$^{-2}$ in the 75~keV~-~2~MeV energy range
for a typical range of GRB models and sky positions (see
Figure~\ref{fig:spectra}). Assuming the reference distance to the
event of $D=410$~Mpc \citep{triggerpaper} this implies an upper limit
on the isotropic equivalent luminosity of $E_{\gamma} < 2 \times
10^{48} \mathrm{erg}
\left(\frac{F_{\gamma}}{10^{-7}\mathrm{erg~cm}^{-2}}\right)
\left(\frac{D}{410 \,\mathrm{Mpc}}\right)^{2}$.  The LIGO
observation corresponds to the energy emitted in gravitational waves
$E_\mathrm{GW} =1.8\pm0.3 \times 10^{54}$~erg. Our SPI-ACS upper limit
constrains the fraction of energy emitted in gamma-rays in the
direction of the observer
$f_{\gamma}~<~10^{-6}~\left(\frac{F_{\gamma}}{10^{-7}~\mathrm{erg~cm}^{-2}}\right)\left(\frac{E_\mathrm{GW}}{1.8
  \times 10^{54}~\mathrm{erg}}\right)^{-1} \left(\frac{D}{410
  \,\mathrm{Mpc}}\right)^{2}$.

\subsection{BH+BH and circumbinary environment}

The analysis of the gravitational wave signal indicates that it was
produced by the coalescence of two black holes
\citep{triggerpaper}. If at least one of the merging black holes was
charged, following the Reissner-Nordstrom formulation, up to 25\% of
the gravitational energy could have been converted into
electromagnetic radiation \citep{zilhao12}. However, it is expected
that the charge of the black hole is spontaneously dissipated 
and is not significant for astrophysical applications. There is no
theoretical work to date predicting electromagnetic emission from the
coalescence of two non-charged back holes in vacuum. Indeed, it is not
possible to create photons in a system with no matter outside of the
gravitational horizon and only gravitational interaction involved,
without invoking effects of quantum gravity, a theory which has not
been developed, yet.

The coalescing black holes may be surrounded by matter, in a form of spherically
symmetric inflow or/and an accretion disk, which can form if the inflow possesses sufficient angular momentum. The accretion
disk can have  high density and large potential energy.  Rapid changes
in the accretion dynamics during binary coalescence may lead to bright
observational signatures \citep{farris12}. Magnetic fields, anchored
in the accretion disk, can cause bright radio emission simultaneous
with the gravitational waves \citep{mosta10}.

While supermassive black holes are often accompanied by substantial
disks, black holes of stellar mass lose the disk created during the
progenitor star collapse on a time scale of the order of
$\tau_\mathrm{disk}\sim 100~\mathrm{s}$ \citep{woosley93}. Sustainable
accretion disks can be expected when a constant inflow of matter is
provided by a companion star: in these cases, the black hole -- star
binary can be a bright and variable X-ray and gamma-ray
source. However, it remains to be established how likely it is to find
a dynamically stable triple system composed of a binary black hole and
an additional companion star.

Isolated stellar-mass black holes or binary black holes are bound to
accrete from the interstellar medium (ISM). This process can be
described as quasi-spherical Bondi-Hoyle accretion \citep{bondi44},
characterized by very low accretion rates $\dot{M}\sim
10^{15}$~g~s$^{-1}$~$\left(\frac{M_H}{65 \times
  \mathrm{M}_\sun}\right)^2~\left(\frac{\rho_\infty}{10^{-24}\mathrm{g~cm}^{-3}}\right)\left(\frac{c_s}{10~\mathrm{km~s^{-1}}}\right)^{-3}$. In
the case of a merger the accretion rate may be enhanced by up to two
orders of magnitude on a one-second time scale \citep{farris09}, but
in the case of a stellar black hole binary accreting from the ISM, the
isotropic peak luminosity can not exceed $L_\mathrm{iso}=100 \times
0.3 \times \dot{M}c^2=2.5\times 10^{37}$~erg~s$^{-1}
\left(\frac{M_H}{65 \,
  \mathrm{M}_\sun}\right)^2~\left(\frac{\rho_\infty}{10^{-24}\mathrm{g~cm}^{-3}}\right)\left(\frac{c_s}{10~\mathrm{km~s^{-1}}}\right)^{-3}$. This
luminosity is almost 17 orders of magnitude lower than the GW
luminosity and more than $\sim$11 orders of magnitude below the
current gamma-ray upper limits.  \cite{agol02} calculated a possible
range of Bondi-Hoyle accretion rates in a Milky Way-like galaxy,
yielding in very rare cases $\dot{M}\sim10^{17}$~g~s$^{-1}$ or peak
luminosity $L_{iso}=3\times10^{39}$~erg~s$^{-1}$, still a factor
$10^9$ below what is observable. The conditions necessary to produce
observable emission may be reached in dense molecular clouds, where
$\left(\frac{\rho_\infty}{10^{-16}\mathrm{g~cm}^{-3}}\right)\left(\frac{c_s}{1~\mathrm{km~s^{-1}}}\right)^{-3}>1$. Therefore,
our upper limit on the hard X-ray burst associated with the merger
disfavors a possibility that the binary was embedded in such a cloud,
unless the emission was very anisotropic.

Recently, different mechanisms to produce the gamma-ray emission in a
black hole binary merger were suggested. For example, a binary black
hole with a very small separation could be formed immediately after
the collapse of a massive star, resulting in a gamma-ray burst
produced nearly simultaneously with a gravitational wave signal
\citep{loeb16}. Alternatively, if an unusually long-lived disk is
present around the black hole binary it could produce bright gamma-ray
signature at the time of the coalescence \citep{perna16}.

\subsection{Alternative possibilities}

\cite{triggerpaper} were able to make use of the gravitational wave data
to constrain the compactness of the merging objects, excluding a
possibility that either of them is a neutron star.

Strange stars are more compact than neutron stars, and their
coalescence can have different gravitational wave signatures
\citep{moraes14}. Very exotic equation of states for strange quark stars would allow
them to reach 6 M$_{\Sun}$ \citep[][not considering
  rotation]{kovacs09}. This is well below the 90\% lower limit inferred for
this event (25 M$_\sun$).

Boson Stars (see \citealt{schunck08} for a review) might reach
arbitrarily high masses, while being only slightly bigger than their
gravitational radius. The existence of these objects requires an
extension of the minimal standard model with a new fundamental scalar
field, responsible for a stable particle. The properties of this field
would determine the macroscopic properties of boson stars. This field
has to be compatible with the non-detection by particle physics
experiments on Earth, cosmological simulations, and models of stellar
evolution. Because of these limitations, the preferred model is
generally a field with minimal coupling to standard model fields. A
boson star consisting of non-charged scalar field can not be directly
involved in any electromagnetic radiation, even in the case of an
energetic coalescence event. On the other hand, the coalescence of
boson stars might have distinct gravitational wave signatures
\citep{palenzuela07}.

Another exotic star kind, Q-stars \citep{bahcal89,bahcal90,miller98}
(where Q here does not stand for quark) can reach 10 or even 100 solar
masses. The existence of these objects was suggested based on finding
a possibility of a peculiar barionic state of matter, without
introducing new matter fields. No predictions on their coalescence
exist to the best of our knowledge.

\section{Conclusions}

We have derived an upper limit on the gamma-ray emission associated
with the gravitational wave event \gwevent\ for the whole localization
region with INTEGRAL. This sets an upper limit on the ratio of the
energy directly released in gamma-rays in the direction of the
observer to the gravitational wave energy E$_\gamma/$E$_\mathrm{GW}<10^{-6}$
(E$_\gamma$ in 75-2000~keV).  This limit excludes the possibility that
the event is associated with substantial gamma-ray radiation, directed
towards the observer.

The LIGO trigger reconstruction favors a binary black hole
scenario. In this case, almost no detectable gamma-ray emission is
expected, unless the binary is surrounded by a very dense gas cloud,
and the emission caused by the enhancement of the accretion rate
during the coalescence is directed towards the observer.

If at least one of the objects is an exotic star (unusually massive quark star, boson
star, Q-star, etc), some electromagnetic emission can not be
excluded. Unfortunately, very little predictions for electromagnetic
signatures of exotic star coalescence are available so far, and our
upper limit provides a constraint for future modeling.

For the first time we have set an upper limit on the gamma-ray
emission associated with a binary black hole merger. This is the
tightest limit that can be set on \gwevent\ with any modern instrument
in the gamma-ray energy range. The emerging possibility of combining
observations of gravitational waves and electromagnetic radiation sets
the beginning of a new era in multi messenger astrophysics.

\section*{Acknowledgements}
\acknowledgments Based on observations with INTEGRAL, an ESA project
with instruments and science data centre funded by ESA member states
(especially the PI countries: Denmark, France, Germany, Italy,
Switzerland, Spain), and with the participation of Russia and the
USA. The SPI-ACS detector system has been provided by MPE
Garching/Germany. We acknowledge the German INTEGRAL support through
DLR grant 50 OG 1101. The Italian INTEGRAL/IBIS team acknowledges the
support of ASI/INAF agreement n. 2016-025-R.0. Some of the results in
this paper have been derived using the \software{HEALPix} \citep{healpix}
package. We are grateful the Fran\c cois Arago Centre at APC for
providing computing resources, and VirtualData from LABEX P2IO for
enabling access to the StratusLab academic cloud. 

\bibliographystyle{apj}
\bibliography{references}

\end{document}